# EMULATING A COMPUTING GRID IN A LOCAL ENVIRONMENT FOR FEATURE EVALUATION

Jananga Kalawana[1], Malith Dilshan[1], Kaveesha Dinamidu[1], Kalana Wijethunga[1,2], Maksim Stortvedt[2] and Indika Perera[1]

[1]Department of Computer Engineering, University of Moratuwa, Bandaranayake Mawatha, 10400, Moratuwa, Sri Lanka
[2]CERN, Esplanade des Particules 1, 1217, Meyrin, Switzerland

## ABSTRACT

*The necessity for complex calculations in high-energy physics and large-scale data analysis has led to the development of computing grids, such as the ALICE computing grid at CERN. These grids outperform traditional supercomputers but present challenges in directly evaluating new features, as changes can disrupt production operations and require comprehensive assessments, entailing significant time investments across all components. This paper proposes a solution to this challenge by introducing a novel approach for emulating a computing grid within a local environment. This emulation, resembling a mini clone of the original computing grid, encompasses its essential components and functionalities. Local environments provide controlled settings for emulating grid components, enabling researchers to evaluate system features without impacting production environments. This investigation contributes to the evolving field of computing grids and distributed systems, offering insights into the emulation of a computing grid in a local environment for feature evaluation.*

## 1. INTRODUCTION

In contemporary computational environments, the significance of computing power cannot be overstated when it comes to executing intricate tasks. While supercomputers provide a substantial amount of computing power as standalone systems, they are constrained by economic factors, limiting access to such power for personal users. Additionally, another significant issue is that supercomputers typically do not allow remote access due to security concerns, thereby hindering their ability to be shared between different organizations. This is where grid computing emerges as a solution. Unlike standalone systems, a grid comprises a network of interconnected computers that may be spatially or globally distributed. Grids offer computing power to users in a convenient, accessible, and reliable manner [1].

One notable example of a grid system is the CERN ALICE [2] computing grid, which provides the massive computational power required for conducting complex calculations in high-energy physics and large-scale data analyses. These grid systems rely on specific site services that provide functionalities such as storage systems, web servers, workload management, and end-user analysis. Among these, workload management encompasses the job submission process, creation, submission, and tracking. However, evaluating modifications to the software responsible for managing these workload services poses a significant challenge. Directly implementing changes in production may result in disruptions and consume considerable time.

The research question addressed by this paper is how to develop a local emulation of a computing grid that enables the reliable and efficient evaluation of new features. The features mentioned in this context refer to new functionalities, optimizations, or enhancements in the grid computing systems' software and infrastructure. These features can include improvements in workload management processes, integration of advanced monitoring tools, security enhancements, or support for new types of computational tasks and workflows. Evaluating these features reliably and efficiently is crucial because they can significantly impact the performance, reliability, and usability of the grid computing system. Testing these features in a local emulation environment allows for thorough assessment and refinement, ensuring that they perform as expected and do not introduce any adverse effects when deployed in the production environment.

Existing computing grid systems compound the challenge addressed by this research paper. The complexity of these grids, coupled with their diverse services and functionalities, makes evaluating newly added features a complex task. This paper proposes a methodology for addressing the research question by presenting a novel approach to emulating a computing grid in a local environment. This approach uses only necessary components to evaluate targeted features while consuming minimal computing resources. We discuss a practical application of this methodology within the implementation: emulating the ALICE grid at CERN in a local environment. The targeted feature, in this case, is a middleware called JAliEn [2], which handles the job submission, tracking, and management processes within the ALICE grid.

The current systematic testing process for evaluating new features in the JAliEn middleware of the ALICE grid involves initial testing on a local test site. If the tests are successful, the version is then promoted to the broader grid, where it undergoes further testing on selected sites with varying setups. The version is only promoted to the full grid after it has successfully operated on these sites for a while. This approach can lead to undesirable outcomes if the newly added features are inconsistent or unstable. This is why the ALICE grid needs the implementation proposed here to evaluate JAliEn before production. There is an existing emulated grid setup referred to as JAliEn Setup or JAliEn Replica for the ALICE grid. However, due to its limitations, such as the lack of automation and the absence of a standard test suite, it is not currently used. Our implementation addresses these issues by applying the proposed methodology to modify and significantly improve the existing setup, adding essential features such as automation and comprehensive testing capabilities. The specific contributions of this paper are:

1. A methodology to emulate a computing grid in a local environment to evaluate targeted features.
2. Improvements made to the JAliEn Setup, providing a more efficient evaluation of JAliEn middleware.
3. Evaluation of the improved JAliEn Setup.

## 2. LITERATURE REVIEW

Computational approaches are employed for problem-solving tasks, yet some complex problems demand significant computational power. Supercomputers offer one solution, but access to their computing power is typically limited to specific organizations rather than individual users.

Utilizing supercomputers often involves single organization ownership and restricted network access, making it challenging for users from other organizations to access their computing power. While remote access is possible, it entails significant complexity due to these restrictions. Consequently, Computing Grids have emerged as hardware and software infrastructures with large-scale resources, facilitating increased accessibility and reliability of computational power. Within the grid architecture, the collective layer responsible for coordinating multiple resources encompasses several essential services, including directory services, scheduling and brokering services, monitoring and diagnostics services, and workload management services [3]. Currently, existing grids include Italy's INFN grid [4], India's Grid GARUDA [5], and CERN-associated WLCG (Worldwide LHC Grid) [6].

The INFN Grid project focuses on developing and deploying grid middleware services to enable users to share computing and storage resources for scientific collaborations. For workload management, INFN Grid utilizes a system called The Workload Management System (WMS), responsible for distributing and managing tasks across grid resources efficiently and effectively. This functionality is achieved through several components, including Computing Elements, Storage Elements, and the Resource Broker consisting of Task Queue and Match Maker components [4].

Another critical use case of computing grids is within the context of high-energy physics experiments, such as those conducted at the Large Hadron Collider (LHC) at CERN, which generates massive amounts of experimental data [7]. Computing grids are employed to handle large-scale data analytics and complex calculations for specific experiments. Each computing grid within the LHC has its own workload management system [8]. For example, the ATLAS [9] experiment computing grid utilizes PanDA[10], consisting of job pilots that retrieve payloads from a central server and execute them. Similarly, the CMS [11] experiment computing grid employs the GlideinWMS[12] workload management system, which creates a virtual pool of job pilots for job submission and execution.

The DIRAC [13] grid infrastructure associated with the LHCb [14] computing grid features a workload management system comprising several components, including a Central Job Management Service, distributed job agents, and job wrappers. Likewise, the ALICE computing grid employs JAliEn as its workload management system, consisting of several components such as central servers, computing elements, and job agents. In all of these systems, jobs are submitted to the computing grids and are subsequently handled by their respective workload management systems, which utilize existing computing resources to execute jobs efficiently and effectively.

In the context of computing grid evaluation, SimGrid [15] is a widely recognized tool for modeling and simulating computing grids. While SimGrid excels in theoretical analysis and high-level modeling of system behavior, it may not capture the intricate details of real-world environments or meet the specific needs of certain systems. In contrast, the proposed methodology utilizes containerized emulation with production images, ensuring that feature evaluations closely mirror the production environment. Additionally, this approach supports integration with CI/CD pipelines and can be fully automated, enabling seamless evaluation workflows that activate with new codebase changes.

Grid'5000 [16] is a distributed, large-scale, and flexible testbed that provides extensive resources for various experiments in Computer Science. It allows researchers to access significant computing power and conduct experiments in a fully customized software environment. However, to evaluate new or modified features within a computing grid using Grid'5000, developers must simulate the grid environment and services on the allocated nodes. This process can be time-consuming and resource-intensive, requiring extensive configuration. In contrast, the containerized approach proposed in this paper addresses these limitations effectively. With minimal computing resources, a local PC environment is sufficient for setting up the emulated grid. The containerized setup can be deployed quickly using a few commands, significantly reducing both setup time and resource requirements. Additionally, the proposed emulated grid can be seamlessly integrated with a CI/CD pipeline or other workflows, as demonstrated in this paper.

[17] discusses building an automated and self-configurable test bed for grid applications. The solution introduced is a framework called the Automated Emulation Framework (AEF), which allows the emulation of grids and automated execution of related experiments in a cluster of workstations. This approach is based on a cluster of virtual machines (VMs), mapping the VMs to hosts, and deploying them. In contrast, the approach from this paper utilizes containers instead of virtual machines. With this container-based setup in a local environment, the emulated grid can run with significantly lower resource consumption compared to using virtual machines.

The emergence of container technologies, exemplified by platforms like Docker and Kubernetes, has revolutionized the deployment and management of applications in modern computing environments. Containers offer lightweight and portable packaging, enabling seamless movement across diverse environments while providing isolation, security, and efficiency. Docker, in particular, stands out for its user-friendly interface and widespread adoption, facilitating agile application development and deployment, especially in cloud environments and microservices architectures. Although Kubernetes offers superior functionality and ease of use compared to Docker Swarm for container orchestration, Docker remains a key player in container virtualization, driving innovation and transformation across various industries. A potential strategy involves leveraging a containerized approach to emulate grid components and locally enumerate the production grid [18, 19].

As in [20], existing container-based emulation techniques, including Mininet [21], Containernet [22], NEMO [23], and openLEON [24], showcase the advantages of using containers for network emulations. Containers can be quickly set up and deployed on a host system, enabling developers to establish their test environments conveniently and efficiently, thus reducing both cost and complexity. Mininet is a network emulator designed for prototyping large networks on devices with constrained resources. Containernet, an extension of Mininet, incorporates Docker containers into network emulation, providing a more comprehensive environment for experiments. While these container-based emulators are effective for network emulation, they are primarily focused on network-specific aspects. In contrast, the proposed approach in this paper leverages container technology to emulate a computing grid within a local environment using minimal resources. Unlike existing network emulators, this methodology is tailored to address the complexities of computing grid systems, offering a localized and resource-efficient solution for feature evaluation.

## 3. METHODOLOGY

The methodology for emulating a grid environment locally requires careful planning and design. This section outlines the key steps involved in setting up the computing grid emulation locally and using it to evaluate the targeted features.

### 3.1. Initial Planning and Design

Emulating a grid in a local environment for feature evaluation entails establishing a system that mirrors the original grid or the production grid, emulating its functionalities. The selection of components and services to emulate and the extent to which they are emulated depends on the intended purpose of the local emulation. For instance, if the aim is to test the entire grid locally, the emulated setup should encompass nearly all components and possibly numerous services. However, if the focus is narrower, such as testing the workload management functionality, then only components and services associated with workload management are prioritized.

Therefore, it is paramount to determine the intended use of the emulated setup and identify which grid components are essential for the local environment. This process involves not only pinpointing crucial components but also analyzing their interactions and dependencies. It is imperative to highlight how these components work together to achieve the expected outcomes from the emulated setup. Once this initial data is gathered, the design and architecture of the emulated setup can be thoughtfully constructed by comparing and evaluating it with the original production computing grid.

## 3.2. Containerized Approach for Component Enumeration

To enumerate grid components separately in the local environment, a suitable approach is needed to deploy or run the components in a significant, isolated environment while providing a means to communicate with each other. As a whole, these components should function as the entire grid system in a scalable and efficient manner. Therefore, implementing components in a monolithic way is not appropriate. While using virtual machines for each component is a viable approach, they are not very lightweight and are less convenient to handle and manage compared to the alternative approach [25].

Considering all factors, the selected approach is to use containers such as Docker or Singularity. These containers can be created with specific scripts to act as grid components in the local environment, allowing them to be started, stopped, and managed conveniently and scalably. Therefore, for the planned design architecture, each grid component can be represented by one or more containers. This approach offers advantages such as clear separation of components, providing an isolated environment, flexibility and convenience in management, and being lightweight compared to virtual machines.

## 3.3. Evaluation of New Features

Evaluating newly added features is crucial, as they typically manifest as code updates to an existing codebase repository. To assess these features, it is necessary to set up the grid in the local environment, execute the grid functionalities, and verify that the results align with expectations. Given that the local emulated grid comprises multiple containers, ensuring proper functionality involves conducting tests separately for each container and evaluating the entire grid's functionality.

While manual tests by developers are an option, they are not ideal as they rely on individual expertise. This underscores the importance of having a unified test suite for the entire system. With a single test suite, developers can effectively ensure that their local grids perform as anticipated. Specifically, the test suite evaluates new versions of features and confirms their proper functionality.

However, even with a local enumeration and a test suite in place, evaluating new features presents another challenge. Setting up the emulated setup, particularly a multi-containerized setup, may be time-consuming, requiring individuals to follow specific steps for setup and evaluation, which is not desirable. The proposed methodology involves integrating a CI/CD pipeline to streamline the emulated setup and testing process. With this approach, there is no need for developers to manually test new versions of features from the beginning, as the test suite used in the CI/CD pipeline automates this process on the emulated setup. Consequently, developers can efficiently evaluate newly added features and understand whether they have disrupted the system.

## 3.4. JAliEn Setup as a Use Case

A practical use case for the proposed methodology can be illustrated through the ALICE computing grid at CERN, which handles complex calculations for the ALICE experiment. The ALICE grid uses JAliEn middleware for job submission, tracking, and management. Modifications to JAliEn can be identified as features in this context, and deploying these modifications often presents challenges. To address this, the JAliEn Setup or JAliEn Replica was introduced, enabling local emulation of the ALICE grid using a setup of five Docker containers. This emulated grid setup facilitates necessary feature evaluation, specifically targeting the JAliEn middleware, ensuring that the workload management functions correctly in a local environment. The following steps outline the application of the proposed methodology in developing the JAliEn Setup.

### 3.4.1. Identify Necessary Grid Components Associated with JAliEn Middleware

The ALICE computing grid comprises several key components associated with the JAliEn middleware: JCentral (Central Servers), Computing Element, Storage Element, Batch Queue, and Worker Nodes. Although the production ALICE grid features multiple sites with numerous Computing Elements, Storage Elements, Batch Queues, and Worker Nodes, the emulated grid focuses on a single instance of each component. This simplification is sufficient for evaluating JAliEn's job submission process in a local environment, optimizing the process, and reducing resource usage.

### 3.4.2. Use Containers to Enumerate Components

Docker containers are highly suitable for representing these components within the context of the ALICE grid's job submission process [26]. Docker Compose can be utilized to orchestrate these containers, configuring their environments to execute necessary functionalities independently. The JAliEn Setup uses the same Docker images as the production ALICE grid components, ensuring a high degree of fidelity. By mounting the containers to the modified JAliEn codebase, jobs can be submitted, tracked, managed, and assessed to evaluate JAliEn.

### 3.4.3. Evaluation of Modified JAliEn

The methodology involves preparing a test suite to evaluate JAliEn. This test suite assesses the containers, submits sample jobs, compares job outputs, and provides results indicating which tests have passed or failed. Developers can review these results to determine whether the modified JAliEn functions as expected or identify any associated failures. Additionally, an automated script integrates the setup and test suite with a CI/CD pipeline. This setup allows for an appropriate workflow that triggers new changes to the JAliEn codebase, automatically starting the container setup, running the test suite, and displaying the results.

## 3.5. Summary

Emulating a complex computing grid like the CERN ALICE grid in a local environment for feature evaluation requires careful consideration of the intended purpose, essential components, and deployment approach. Utilizing containerization technologies such as Docker or Singularity offers a lightweight and scalable solution for emulating grid components separately while ensuring efficient communication between them. By adopting a multi-containerized approach, developers can accurately test the functionality of individual components as well as the entire grid system. Furthermore, the implementation of a single comprehensive test suite facilitates the thorough evaluation of newly added features, enhancing the efficiency and reliability of the testing process. With automation integrated into the CI/CD pipeline to streamline setup and testing procedures, developers can expedite the evaluation process and promptly identify any issues or discrepancies. This ensures the smooth operation of the grid system in both local and production environments.

# 4. IMPLEMENTATION

The implementation of the JAliEn Setup demonstrates a practical application of the proposed methodology. This section details the steps taken to emulate the ALICE grid locally, ensuring reliable feature evaluation before deployment to production.

## 4.1. ALICE Grid, JAliEn and JAliEn Setup

The ALICE experiment is one of the four major experiments at CERN and is dedicated to heavy ion physics at the Large Hadron Collider (LHC). The ALICE computing grid handles this research's massive and complex calculations. JAliEn is the middleware used in the ALICE grid for job creation, submission, execution, tracking, and management. Users of the ALICE grid can submit computational jobs to the grid and receive output results for those submitted jobs. JAliEn oversees this entire process. However, frequent modifications to the JAliEn codebase present a challenge when deploying new changes to production. If the new JAliEn version fails during site-by-site deployment, it can take a significant amount of time and resources to complete the deployment.

As proposed in this paper, by emulating the ALICE grid in a local environment to evaluate JAliEn, developers can carry out those evaluations locally, detect potential issues, and stabilize JAliEn before starting the deployment process. The JAliEn Setup (also known as JAliEn Replica) was developed to facilitate this. The initial version of the JAliEn Setup allowed the ALICE grid to be emulated locally in a containerized setup, but it only covered part of the proposed methodology and had several limitations. These limitations included the absence of an accepted test suite and the need for developers to manually perform evaluations each time. The implementation associated with this research aims to improve the existing JAliEn Setup by addressing these limitations and providing a more robust and automated solution for local emulation and testing.

## 4.2. The Architecture of the JAliEn Setup

The targeted feature to evaluate here is the job submission flow of the ALICE grid, which is managed by the JAliEn middleware. The JAliEn Setup was developed to emulate the ALICE grid, enabling the job submission flow to be run and tested locally.

There are five main components in the ALICE grid associated with JAliEn. Similarly, as shown in the figure 1, the JAliEn Setup also consists of five main components:

- JCentral - Central service
- CE - Computing element
- Schedd - Scheduler

- Worker - Worker node
- SE - Storage element

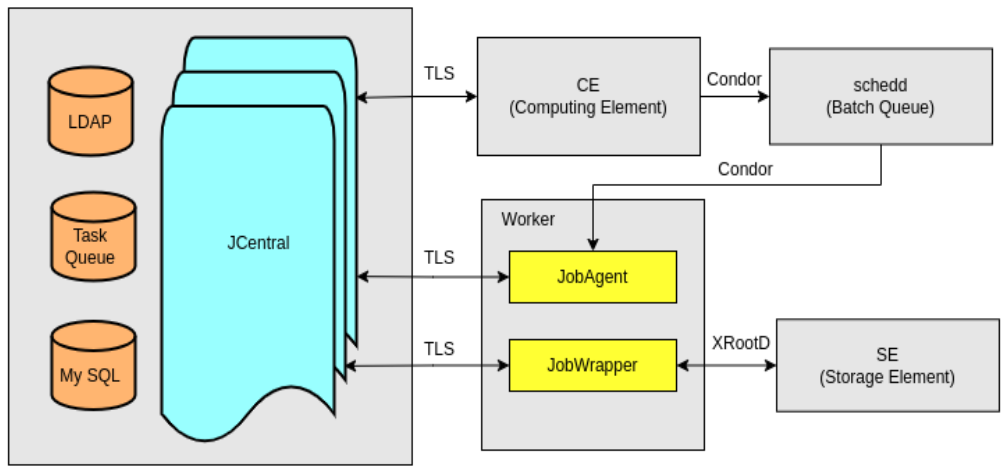

Figure 1. Architecture diagram of the emulated setup

Each of these components is housed in a dedicated container, and together, these five containers emulate the ALICE grid. The modified JAliEn codebase is mounted for the JCentral, CE, and Worker node containers. This container setup can be used to evaluate the job submission flow functionalities of JAliEn.

### 4.2.1. JCentral

In JAliEn, "JCentral" refers to the central services that manage the entire system. It coordinates tasks such as authentication, authorization, and database services. The component handling user requests and requests from other components likely interacts with JCentral for various operations, ensuring smooth communication and functionality across the system [27]. This component utilizes the Ubuntu 18 image.

### 4.2.2. Computing Element

This component is responsible for announcing the available resources to JCentral and for creating job agents. Job agents are special programs that run on worker nodes. They request jobs to execute from JCentral [28] and start jobs on workers. This component utilizes the AlmaLinux 9 image.

### 4.2.3. Scheduler

The scheduler component in the grid uses HTCondor as the Batch Queue [29], utilizing the HTCondor Scheduler image. The Computing Element (CE) creates a startup script to create a job agent and submits it to the batch queue [28]. The scheduler then assigns the job to a worker node.

### 4.2.4. Worker

On the worker node, the job agent starts a job by creating a separate process called JobWrapper. The JobWrapper process is responsible for downloading input files from the Storage Element (SE), executing the job, and uploading output files back to the SE [28]. The worker node utilizes a CentOS 7 image for its operation.

**4.2.5. Storage Element**

The storage elements of the production grid are distributed among different sites to form a distributed storage system. This distributed storage setup allows for efficient and scalable storage of data across the grid.

Each storage element utilizes an XRootD [30] image. XRootD is a software framework for data access, storage, and management. It provides a scalable and high-performance solution for distributed data access, making it well-suited for use in grid computing environments.

The storage elements store both user data and system data necessary for the operation of the grid [2]. They play a crucial role in ensuring data availability, reliability, and performance across the grid infrastructure.

## 4.3. Flow of the Emulated Setup

The job submission flow of the emulated grid is identical to the job submission flow of the production ALICE grid. When the JAliEn Setup is started using Docker Compose, it initializes the containers, beginning with the JCentral container. JCentral initializes the database and other necessary configurations, starts a web server, and listens for incoming requests from other container components. The Computing Element periodically checks for available jobs by communicating with JCentral.

Jobs can be submitted to JCentral as scripts via shell commands. Upon receiving a job submission, JCentral updates its databases. When the Computing Element checks for available jobs, it identifies the new job, creates a job agent startup script, and submits that script to the Batch Queue (Scheduler Container). The Scheduler Container, which contains the HTCondor scheduler image, starts job agents on the Worker Node according to the submitted startup scripts. The job agent runs on the Worker Node and requests job execution details from JCentral. If a job is available, the job agent starts a separate process called Job Wrapper on the Worker Node. The Job Wrapper process executes the job, updates the job status, and stores the results in the Storage Element. Users can check the job status and access the final output results via shell commands to JCentral.

When evaluating the functionality of JAliEn, all aspects of this job submission flow must be covered. This means ensuring that all containers are healthy and functioning properly, submitting a job or jobs to verify that the process is carried out correctly, and finally checking the final outputs. If there is a failure at any point, it should be identified along with a probable reason for the failure.

## 4.4. Test Suite for JAliEn Setup

To implement the test suite, Bash scripts were chosen as JAliEn Setup itself uses Bash scripts for some of its necessary tasks. However, the choice of implementation technology ultimately depends on the context, and it would be perfectly fine to utilize any technology as long as it is compatible with the emulated grid. The test suite is developed such that, with a single command, developers can run it. For customizability, some flags are introduced with the starting command, allowing developers to run either the full test suite or a part of it.

First, the test suite checks whether the host system, where the emulated grid is set up, has all the necessary libraries installed and other environmental configurations. Then, the test suite evaluates each container component separately by checking whether the necessary libraries have been installed and their environmental configurations. Most importantly, the test suite focuses on the job submission process with JAliEn. As this job submission process is carried out, it generates certain logs and files in separate containers, so the job submission process can be evaluated by checking the content of those logs and files. The test suite submits jobs and checks for those logs and files. Additionally, the test suite periodically checks the status of the job to see whether it is done or failed. When the job is done, the test suite checks the job result or the outputs to ensure they are as expected.

The implemented test suite has three types of tests: critical, warning, and minor. Critical tests relate to critical aspects of the flow and components. If such a test fails, the test suite run fails, showing the respective error messages. In the case of warning and minor tests, their failure shows the message and probable reason for failure but does not stop the test suite execution. After the test run finishes, developers can review failed tests to identify any issues with the new version of JAliEn. This test suite ensures all developers evaluate JAliEn in a standard and agreed way, ensuring new JAliEn modifications are stable and reliable. Figure 2 and 3 show an output generated while running the test suite and the test summary output after the test run, respectively.

Figure 2. Output during test execution

Figure 3. Test suite summary

## 4.5. CI/CD Pipeline

With the JAliEn Setup and the implemented test suite, JAliEn can be evaluated, but it must be done manually, which can be time-consuming. To address this, an automated script setup was developed using Bash. This allows the JAliEn Setup to be started and the test suite to be run with a single customizable command. The selection of technology for the automated scripts depends on the context and must be compatible with the emulated grid and the test suite.

Furthermore, the emulated JAliEn Setup, test suite, and automated script setup are integrated into a CI/CD pipeline in the form of a GitHub workflow, as shown in Figure 4. When a new update is pushed to the JAliEn codebase repository, it triggers the workflow, starts the JAliEn Setup, runs the test suite, and shows the test outputs. Developers can then check the outputs to ensure the updates are reliable. This workflow ensures that all changes to the codebase are properly evaluated before being sent to production.

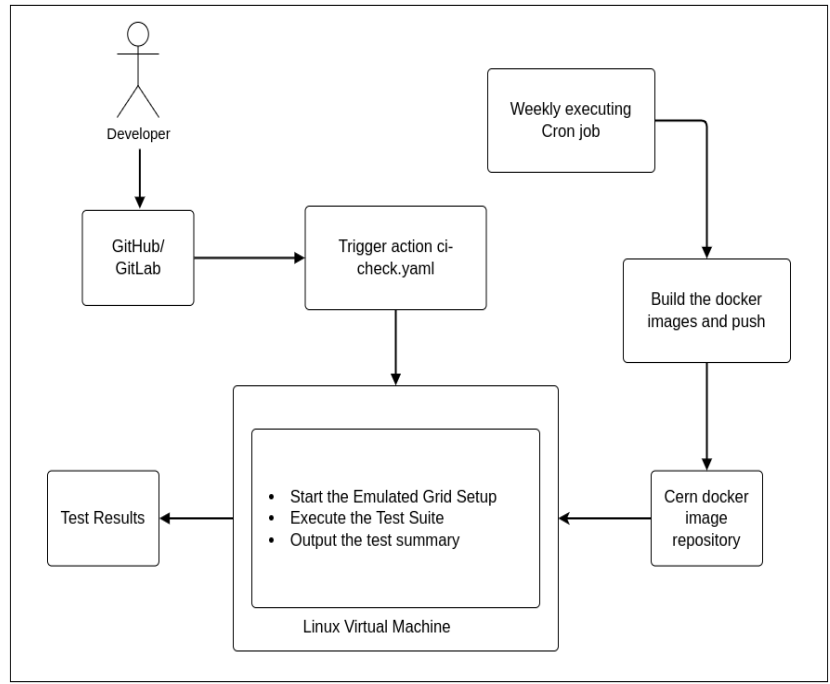

Figure 4. CI/CD pipeline structure

In addition to this, there is another periodically running workflow set up as a cron job to run weekly. This workflow builds the Docker images for container components and uploads them to the CERN Docker container registry. Therefore, when the JAliEn Setup is used, local environments do not need to build the images; they can simply pull them from the registry and proceed. This significantly reduces the time required to start the JAliEn Setup.

## 5. RESULTS AND EVALUATION

Currently, the JAliEn Setup is capable of emulating the ALICE grid in a local environment using a five-container setup. This emulated grid, along with the specifically developed test suite and automated CI/CD pipeline, allows for a thorough evaluation of the job submission feature via the JAliEn middleware. The following section discusses the overall results and evaluation of the improved JAliEn Setup.

### 5.1. Results

The modified JAliEn Setup successfully emulated the ALICE grid in the local environment, enabling job submission and tracking using the emulated setup. By using the same Docker images as the production environment for the emulated grid components, the evaluation of JAliEn occurs in an environment identical to the production setup.

The test suite covers all critical aspects of the JAliEn middleware related to job submission within the emulated grid. It first verifies the individual components separately and then tests the entire job submission flow. The test suite is capable of detecting failures in job submission functionality, providing error messages or warnings along with probable reasons for the failures. Figures 2 and 3 show sample outputs from running the test suite and the test suite summary. This capability significantly aids the development process by identifying and fixing issues efficiently.

Moreover, these procedures are integrated into CI/CD pipelines within the respective repositories as workflows. This integration ensures that the grid is emulated, tests are conducted, and results are displayed whenever a new change is introduced. Consequently, developers can verify the reliability of new changes before pushing them to production, thereby reducing failure rates in the production environment.

## 5.2. Qualitative Evaluation

Implementing the modified JAliEn Setup from the proposed methodology introduces several advantages over existing systems and procedures.

**Resource Efficiency:** The JAliEn Setup uses minimal resources by employing only essential grid components in the form of Docker containers. This contrasts with traditional grid simulations that often require extensive resources and infrastructure.

**Failure Rate Reduction:** The methodology significantly reduces failure rates in the production environment by emulating the grid environment locally and conducting rigorous testing before deploying it to production.

**Convenient Usage:** The JAliEn Setup can be fully automated using the provided scripts or through the CI/CD pipeline. This convenience allows developers to set up the emulated grid and evaluate JAliEn with minimal effort.

The modified JAliEn Setup (v2) has the following new capabilities compared to the previous JAliEn Setup (v1),

- **Comprehensive Testing:** The development of a detailed test suite allows for a thorough evaluation of the job submission process. This level of testing, from individual components to the entire workflow, enhances the system's reliability.
- **Automation and Integration:** The ability to execute the entire process manually or through automated scripts streamlines operations. Additionally, integrating these procedures into CI/CD pipelines ensures continuous testing and validation of new changes, which is a significant improvement over manual deployment and testing processes.

Table 1 provides a summary of the qualitative comparison between JAliEn Setup v1 and JAliEn Setup v2.

Table 1. Feature comparison

| Feature | JAliEn Setup v1 | JAliEn Setup v2 |
|---------|-----------------|-----------------|
| Build docker images | Yes | Yes |
| Manually start the emulated setup | Yes | Yes |
| Test suite to assess functionalities | No | Yes |
| Automated scripts for setup and running tests | No | Yes |
| CI/CD pipeline | No | Yes |

## 5.3. Quantitative Evaluation

Since the proposed methodology cannot be directly evaluated quantitatively, the modified JAliEn Setup (v2) is evaluated against the existing JAliEn setup (v1) using execution time as the metric. The setups were run on an Acer Predator Helios 300 PC equipped with Intel(R) Core™ i7-6550U (1.80GHz) 8-core processors running Ubuntu 22.04.4 LTS Operating System. With JAliEn Setup v1, the grid can be emulated using the production images, but feature evaluation must be done manually. In contrast, JAliEn Setup v2 not only emulates the grid but also provides means to evaluate the features automatically.

Times taken for building Docker images, starting the emulated grid, and evaluating the job submission flow have been collected using both versions of the JAliEn setup. The average times for each step are listed in table 2.

Table 2. Time comparison between JAliEn Setup v1 vs v2

| | v1 (minutes) | v2 (minutes) |
|---|---|---|
| Built docker images | 86 | 89 |
| Start the emulated grid | 10 | 6 |
| Evaluate the job submission flow | 42 | 13 |

As shown in table 2, except for building Docker images, the times taken for the other two steps are significantly less in JAliEn Setup v2. Additionally, images are built periodically each week and pushed to a container registry, so there is no need to build them from scratch unless developers make changes to the images themselves and need to test those changes.

This quantitative evaluation demonstrates the efficiency and effectiveness of JAliEn Setup v2 in terms of reducing setup and evaluation times, thereby facilitating quicker and more reliable testing of new updates to JAliEn.

## 5.4. Limitations

The developed JAliEn Setup effectively evaluates the job submission functionality using the test suite and CI/CD pipeline within the local emulated grid. However, the grid components are containerized in the local setup, whereas in the production grid, these are separate, distributed components. This difference may lead to unpredictable issues such as communication breakdowns or node failures in the production environment that the emulated setup might not fully capture. While the JAliEn Setup test suite can identify many issues, certain edge cases that might occur in production, where JAliEn could behave unexpectedly, may not be detected. Such issues can only be fully identified and resolved when running in the actual production environment, beyond the scope of the emulated setup.

## 6. CONCLUSION

The proposed methodology for emulating a computing grid in a local environment for feature evaluation has been successfully applied in implementing an emulated grid for the ALICE grid and using it to evaluate the job submission feature by the JAliEn middleware. This emulated grid, called JAliEn Setup or JAliEn Replica, has proven its worth in testing and evaluating modified versions of JAliEn. The multi-containerized nature of the JAliEn Setup, with only essential grid components, ensures it uses just enough computing resources for the intended task. The modifications made to the JAliEn Setup during the implementation phase ensure that the emulated setup is up-to-date and compatible with the production grid. Furthermore, developing a test suite to verify JAliEn functionality has proven to be quite useful for standardizing testing for each developer.

Integrating the emulated grid and test suite with a CI/CD pipeline adds automation, further ensuring that only reliable and stable updates are pushed to the JAliEn in production. Overall, this entire JAliEn setup system evaluates the job submission feature of the ALICE grid carried out by the JAliEn middleware, ensuring its reliability and stability before deployment on production sites. Consequently, when new features are deployed, there will be fewer failures in production, saving both time and resources.

Considering the proposed methodology and implementation, this approach can be applied to other computing grids to emulate a grid locally for evaluating specific features. Since many computing grids have core elements that function in a similarly approximate way, such as Computing Elements and Batch Queues, this implementation of JAliEn Setup can serve as a reference for applying the proposed methodology and implementing an emulated grid for other computing grids as well.

## ACKNOWLEDGEMENTS

We would like to extend our deepest gratitude to our university and the CERN organization for providing us with the invaluable opportunity to contribute to this project. We are especially thankful to our supervisors, both from the university and CERN, who consistently encouraged us to push beyond our limits. Lastly, we are profoundly grateful for the gift of free education, which has made all of this possible.

## REFERENCES

[1]     C. E. A. Karnow, "The Grid: Blueprint for a New Computing Infrastructure edited by Ian Foster and Carl Kesselman Morgan Kaufman, San Francisco, CA, U.S.A., 1998. 677 pp. ISBN: 1558-604-758.," Leonardo, vol. 32, no. 4, pp. 331–332, Aug. 1999, doi: 10.1162/leon.1999.32.4.331.

[2]     M. M. Pedreira, C. Grigoras, and V. Yurchenko, "JAliEn: the new ALICE high-performance and high-scalability Grid framework," EPJ Web of Conferences, vol. 214, p. 03037, Jan. 2019, doi: 10.1051/epjconf/201921403037.

[3]     I. Foster, C. Kesselman, and S. Tuecke, "The Anatomy of the Grid: Enabling Scalable Virtual Organizations," The International Journal of High Performance Computing Applications, vol. 15, no. 3, pp. 200–222, Aug. 2001, doi: 10.1177/109434200101500302.

[4]     F. Donno, L. Gaido, A. Ghiselli, M. Mazzucato, F. Prelz, and M. Sgaravatto, "INFN GRID DATAGRID Prototype 1," Jan. 2002, doi: 10.15161/oar.it/1448983015.84.

[5]     B. B. P. Rao, S. Ramakrishnan, M. R. R. Gopalan, C. Subrata, N. Mangala, and R. Sridharan, "e-Infrastructures in IT: A case study on Indian national grid computing initiative – GARUDA," Computer Science - Research and Development, vol. 23, no. 3–4, pp. 283–290, May 2009, doi: 10.1007/s00450-009-0079-3.

[6]  D. Bonacorsi and T. Ferrari, "WLCG Service Challenges and Tiered architecture in the LHC era," in Springer eBooks, 2007, pp. 365–368, doi: 10.1007/978-88-470-0530-368.

[7]  J. F. Molina, A. Forti, M. Girone, and A. Sciaba, "Operating the Worldwide LHC Computing Grid: current and future challenges," Journal of Physics Conference Series, vol. 513, no. 6, p. 062044, Jun. 2014, doi: 10.1088/1742-6596/513/6/062044.

[8]  A. Sciabà, "Critical services in the LHC computing," Journal of Physics. Conference Series, vol. 219, no. 6, p. 062025, Apr. 2010, doi: 10.1088/1742-6596/219/6/062025.

[9]  J. Andreeva, S. Campana, F. Fanzago, and J. Herrala, "High-Energy Physics on the Grid: the ATLAS and CMS Experience," Journal of Grid Computing, vol. 6, no. 1, pp. 3–13, Sep. 2007, doi: 10.1007/s10723-007-9087-3.

[10] T. Maeno et al., "Overview of ATLAS PanDA Workload Management," Journal of Physics. Conference Series, vol. 331, no. 7, p. 072024, Dec. 2011, doi: 10.1088/1742-6596/331/7/072024.

[11] C. Collaboration et al., "The CMS experiment at the CERN LHC," Journal of Instrumentation, vol. 3, no. 08, p. S08004, Aug. 2008, doi: 10.1088/1748-0221/3/08/s08004.

[12] I. Sfiligoi, "glideinWMS—a generic pilot-based workload management system," Journal of Physics Conference Series, vol. 119, no. 6, p. 062044, Jul. 2008, doi: 10.1088/1742-6596/119/6/062044.

[13] F. Stagni and P. Charpentier, "The LHCb DIRAC-based production and data management operations systems," Journal of Physics. Conference Series, vol. 368, p. 012010, Jun. 2012, doi: 10.1088/1742-6596/368/1/012010.

[14] J. P. Baud et al., "The LHCb Data Management System," Journal of Physics. Conference Series, vol. 396, no. 3, p. 032023, Dec. 2012, doi: 10.1088/1742-6596/396/3/032023.

[15] H. Casanova, A. Legrand, and M. Quinson, "SimGrid: A Generic Framework for Large-Scale Distributed Experiments," Jan. 2008, doi: 10.1109/uksim.2008.28.

[16] R. Bolze et al., "Grid'5000: A Large Scale And Highly Reconfigurable Experimental Grid Testbed," The International Journal of High Performance Computing Applications, vol. 20, no. 4, pp. 481–494, Nov. 2006, doi: 10.1177/1094342006070078.

[17] R. N. Calheiros, R. Buyya, and C. a. F. De Rose, "Building an automated and self-configurable emulation testbed for grid applications," Software, Practice Experience/Software, Practice and Experience, vol. 40, no. 5, pp. 405–429, Mar. 2010, doi: 10.1002/spe.964.

[18] E. Casalicchio and S. Iannucci, "The state-of-the-art in container technologies: Application, orchestration and security," Concurrency and Computation, vol. 32, no. 17, Jan. 2020, doi: 10.1002/cpe.5668.

[19] V. G. Da Silva, M. Kirikova, and G. Alksnis, "Containers for Virtualization: An Overview," Applied Computer Systems, vol. 23, no. 1, pp. 21–27, May 2018, doi: 10.2478/acss-2018-0003.

[20] J. Gomez, E. F. Kfoury, J. Crichigno, and G. Srivastava, "A survey on network simulators, emulators, and testbeds used for research and education," Computer Networks, vol. 237, p. 110054, Dec. 2023, doi: 10.1016/j.comnet.2023.110054.

[21] B. Lantz, B. Heller, and N. McKeown, "A network in a laptop," Oct. 2010, doi: 10.1145/1868447.1868466.

[22] M. Peuster, J. Kampmeyer, and H. Karl, "Containernet 2.0: A Rapid Prototyping Platform for Hybrid Service Function Chains," Jun. 2018, doi: 10.1109/netsoft.2018.8459905.

[23] L. Veltri, L. Davoli, R. Pecori, A. Vannucci, and F. Zanichelli, "NEMO: A flexible and highly scalable network EMulatOr," SoftwareX, vol. 10, p. 100248, Jul. 2019, doi: 10.1016/j.softx.2019.100248.

[24] C. Fiandrino, A. B. Pizarro, P. J. Mateo, C. A. Ramiro, N. Ludant, and J. Widmer, "openLEON: An end-to-end emulation platform from the edge data center to the mobile user," Computer Communications, 2019.

[25] M. M. Storetvedt, M. Litmaath, L. Betev, H. Helstrup, K. F. Hetland, and B. Kileng, "Grid services in a box: container management in ALICE," EPJ Web of Conferences, vol. 214, p. 07018, Jan. 2019, doi: 10.1051/epjconf/201921407018.

[26] M. M. Storetvedt, L. Betev, H. Helstrup, K. F. Hetland, and B. Kileng, "Running ALICE Grid Jobs in Containers A new approach to job execution for the next generation ALICE Grid framework," EPJ Web of Conferences, vol. 245, p. 07052, Jan. 2020, doi: 10.1051/epjconf/202024507052.

[27] A. G. Grigoras, C. Grigoras, and V. Yurchenko, "Running ALICE Grid Jobs in Containers A new approach to job execution for the next generation ALICE Grid framework," Journal of Physics. Conference Series, vol. 1525, no. 1, p. 012034, May 2020, doi: 10.1088/1742-6596/1525/1/012034.

[28]  D. Thain, T. Tannenbaum, and M. Livny, "Distributed computing in practice: the Condor experience," Concurrency and Computation: Practice and Experience, vol. 17, no. 2–4, pp. 323–356, Feb. 2005, doi: 10.1002/cpe.938.

[29]  B. Shirazi, S. V. T. Beek, M. Younis, and H. Qi, "WFCS: A Cluster-Based Framework for Dependable Computing," in Lecture Notes in Computer Science, Berlin, Heidelberg: Springer Berlin Heidelberg, 2000, pp. 1095–1104.

[30]  F. Dechouniotis, D. Kyriazis, and G. M. Poulios, "A framework for dependability assurance in service-oriented systems," IEEE Transactions on Systems, Man, and Cybernetics: Systems, vol. 47, no. 5, pp. 735–746, Dec. 2016, doi: 10.1109/tsmc.2016.2635020.

## AUTHORS

**Jananga Kalawana** is an enthusiastic Computer Science & Engineering undergraduate with a strong interest in grid computing, automation, and problem-solving. Dedicated to contributing to both society and research by bringing innovative ideas to the table and developing efficient solutions.

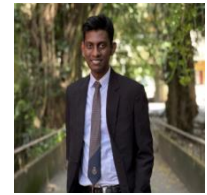

**Malith Dilshan** is a passionate Computer Science & Engineering undergraduate with a keen interest in automation. He is dedicated to contributing to the world by developing solutions that enhance the quality of life for people everywhere.

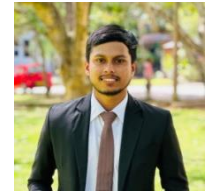

**Kaveesha Dinamidu**  is a Computer Science & Engineering undergraduate with a strong passion for cloud computing and distributed systems. Driven by a desire to create innovative solutions, Kaveesha aims to contribute to a better world by harnessing the power of technology to improve people's lives.

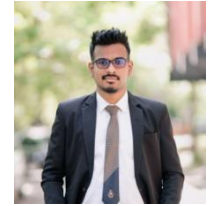

**Kalana Wijethunga** is a skilled software engineer at Canava with 3 years of professional experience adept at navigating complex projects in banking, identity management, and high-performance computing. Worked at CERN, the European Organization for Nuclear Research, Proficient in collaborating with international teams excelling in problem analysis and implementing scalable solutions.

**Maxim Stortvedt** is a proficient software engineer at CERN, the European Organization for Nuclear Research.

**Indika Perera** is a Professor in Computer Science and Engineering at University of Moratuwa.